\documentclass[prd,twocolumn,showpacs,preprintnumbers,nofootinbib]{revtex4}
\usepackage{amsfonts,amsmath,amssymb}
\usepackage{graphicx}
\usepackage{color}
\usepackage{natbib}
\usepackage{siunitx}

\usepackage[plainpages=false, colorlinks=true, anchorcolor=blue, linkcolor=blue, citecolor=blue, bookmarks=false]{hyperref}
\pdfoutput=1
\newcommand{\rthis}[1]{\textcolor{black}{#1}}

\usepackage{natbib}

\begin{document}


\title{Non-Gaussian Error Distributions of Galactic Rotation Speed Measurements}
\author{Ashwani \surname{Rajan}$^1$}  \altaffiliation{E-mail: ashwani.rajan135@gmail.com}
\author{Shantanu  \surname{Desai}$^2$} \altaffiliation{E-mail: shntn05@gmail.com}

\affiliation{$^{1}$Department of Physics, Indian Institute of Technology, Guwahati, Assam-781039, India}
\affiliation{$^{2}$Department of Physics, Indian Institute of Technology, Hyderabad, Telangana-502285, India}

\begin{abstract}
We construct the error distributions for the  galactic rotation speed  ($\Theta_0$)  using 137 data points from measurements compiled in De Grijs et al.\cite{Grijs}, with all observations normalized to the  galactocentric distance of 8.3 kpc.  We  then checked (using the same procedures as in works by Ratra et al) if the errors constructed using the weighted mean and the median as the estimate, obey Gaussian statistics. We find using both these estimates that they have much wider tails than a Gaussian distribution. We also tried to fit the data to three other distributions: Cauchy, double-exponential, and Students-t. The best fit is obtained using the Students-$t$ distribution for $n=2$ using the median value as the central estimate, corresponding to a $p$-value of 0.1. 
We also calculate the median value of $\Theta_0$ using all the data as well as using the median of each set of measurements based on the tracer population used.  \rthis{Because of the non-gaussianity of the residuals, we point out that the subgroup median value, given by  $\Theta_{med}=219.65$ km/sec should be used as the central estimate for $\Theta_0$.} 

\end{abstract}

\maketitle

\section{Introduction}
Recently, de Grijs and Bono~\cite{Grijs} (hereafter G17), compiled a list of 162 galactic rotation speed measurements (denoted as $\Theta_0$) using data from all the published literature starting from 1927  right up to 2017. Two main goals  of this meta-analysis  was to look for evidence for publication bias and to check how close  is the central estimate from all these measurements to the IAU recommended value of  $\Theta_0=220$ km/sec~\cite{Kerr}. Although, no evidence for such a bias was seen, G17 found evidence for systematic biases in the measurements of $\Theta_0$ between the different tracer populations. The estimated value for Galactic rotation speed obtained in G17 using all the post-1985  measurements is given by $\Theta_0= 225 \pm 3 (\rm{stat}) \pm 10 (\rm{syst})$   km/sec, after positing a galactocentric distance of 8.3 kpc.

In the last decade, Ratra and collaborators  have used a variety of astrophysical datasets
to test  the non-Gaussianity of the error distributions from these measurements. The datasets they explored for this purpose include measurements of $H_0$~\cite{Ratra03}, Lithium-7 measurements~\cite{Ratra15} (see also ~\cite{Zhang}), distance to LMC ~\cite{RatraLMC}, distance to galactic center~\cite{RatraGC}. Evidence for non-Gaussian errors has also been found in HST Key project data~\cite{Singh}.
For each of these datasets, they have fitted the data to a variety of probability distributions.  \rthis{For all of these studies, they have found the error distributions to be non-Gaussian. As a consequence they have argued that median statistics should be used for central estimates of these parameters instead of the weighted mean.}
Median statistics  has therefore  been used to obtain central estimates of $H_0$~\cite{Gott,Chen,Bethapudi}, $G$~\cite{Bethapudi}, mean matter density~\cite{Chen03} and other cosmological parameters~\cite{Crandall}.

\rthis{Inspired by these works,  we revisit the issue of getting the best estimate of $\Theta_0$
from the catalog compiled in G17. The first step in this analysis is to check for non-Gaussianity of the residuals. The importance of checking for  non-Gaussianity of the measurement errors for a large suite of astrophysical measurements has been stressed in a number of works in astrophysics ~\citep{Gott,Ratra03,Ratra15,RatraGC,RatraLMC,Singh,Zhang}. Most recently, its importance in other fields such as nuclear and particle physics, medicine and toxicology has also been elucidated  by Bailey~\cite{Bailey}.}

\rthis{It is common practice to assume that datasets are Gaussian. However, this is not always the case. By carrying out Gaussianity tests on error distributions of measurements, several insights can be obtained. One application is the possibility of reduced significance of discrepancies between observed and expected values.
Usually, when a central  estimate  for a physical quantity is needed, one typically calculates a weighted average of all the measurements. One assumption implicitly made herein is that the weighted mean error distributions have a Gaussian distribution. If this is not the case, then one cannot directly use weighted mean estimates or $\chi^2$ analysis for parameter estimation. One then needs to resort to median statistics~\citep{Gott,Bethapudi}, which does not invoke the measurement errors and is not affected by the non-Gaussianity~\cite{RatraGC}. If the residuals are non-Gaussian, one possibility is that the errors are underestimated and there are additional unaccounted systematic errors , which could be ``known unknowns'' or ``unknown unknowns''.  Another possible reason could be due to  outliers, which may arise due to egregious measurements. These outliers could potentially bias any estimates. 
Conversely,  if the tails in a distribution are narrower than a Gaussian, it implies that the different measurements are correlated.
For this reason a large number of studies in astrophysics and other fields have investigated and found evidence for non-Gaussianity for a diverse suite of measurements.}

\rthis{The galactocentric velocity  is of tremendous importance in both galactic astrophysics as well as cosmology, and hence in order to obtain a robust estimate of its central value, one needs to check for non-Gaussianity of errors.}

The outline of this paper is as follows. The dataset used for our analysis is described in Sect.~\ref{sec:dataset}. Our analysis procedure and results are described in Sect.~\ref{sec:analysis}. The corresponding analysis on each sub-group of measurements is discussed in Sect.~\ref{sec:subsamples}. We conclude in Sect.~\ref{sec:conclusions}.

\section{Dataset Used}
\label{sec:dataset}
We briefly review the data on the galactic rotation speed measurements compiled by G15. More details can be found in their paper~\cite{Grijs}. The main goal of their paper (intended as a follow-up to a series of papers~\cite{Grijs14a,Grijs14b,Grijs15,Grijs16} looking for similar effects in other observables) was to undertake a meta-analysis of all the measurements of $\Theta_0$ from published literature in order to look for intrinsic differences between the different categories of measurements of $\Theta_0$. They also wanted to see if there is evidence for publication bias or ``bandwagon'' effect. 

The previous  comprehensive review of literature on galactic rotation velocities was carried out by Kerr and Lynden-Bell~\cite{Kerr},
which explained the reasoning behind the IAU recommended value of 220 km/sec at the solar circle. G17 searched the NASA/Astrophysics  Data System (ADS) by looking for articles referring to the Milky Way and using one of the following keywords in the abstract search: `rotation curve', `kinematics' (including its variants), `dynamics', and 'Oort'. They found a total of 9,690 articles starting from Oort's original papers in 1927~\cite{Oorta,Oortb} until the end of June 2017. They  data mined all these papers looking for new values of Galactic rotation constants.  These papers either provided a direct measurements of the  galactic rotation speed  or the Oort constants $A$ and $B$~\cite{Oorta,Oortb}, from which 
$\Theta_0$ is given by $(A-B)R_0$, where $R_0$ is the Galactocentric distance. Since majority of the $\Theta_0$ measurements hinge on the determination of $R_0$, G17 homogenized all measurements of $\Theta_0$ to a common value of $R_0=8.3$ kpc, based on recommendations from their previous set of studies~\cite{Grijs16}. 

In all, they found a total of 162 measurements.
These consist of seven different types of stellar population tracers. All these measurements along with their statistical uncertainties   have been uploaded on the internet.~\footnote{See \url{http://astro-expat.info/Data/pubbias.html} for a compilation of all these measurements.}  We note that no systematic errors have been included in the analysis. \rthis{In addition to these measurements compiled by G17, we used two additional measurements compiled by  Salucci et al~\cite{Salucci10,Salucci13}, which are not listed in G17. In one of them~\cite{Salucci10}, $\Theta_0/R_0$ has been estimated to be $30.3 \pm 0.9$ km/sec/kpc. In Ref.~\cite{Salucci10}, $\Theta_0$ has been estimated to be $239 \pm 7$ km/sec. We also found that one measurement by Glushkova et al~\cite{Glushkova} was incorrectly tabulated on the website. At the time of writing, the website reported a measurement of 277  $\pm$ 3 km/sec at a distance of 7.4 kpc. However, this is a typographical error on the website and the correct measurement reported in the paper is 204 $\pm$ 15 km/s. For our analysis, we used the correct measurement reported in the paper.}

Out of these 164 measurements, we omitted 26 for which no errors were provided. \rthis{ We also left out the measurement in Ref.~\cite{Evans},  corresponding to a value of 198 km/sec at 10~kpc. This value corresponds to a 23.5$\sigma$ discrepancy compared to the weighted mean value. One possible reason for this low value of the rotation speed~\cite{Evans}  is because of the simplified \emph{ansatz} they used for the Galactic gravitational potential, viz. a spherical power law. }
We also normalized  all the remaining 137  measurements of $\Theta_0$ and their associated errors, which are degenerate with galactocentric distance to a $R_0$ value of 8.3 kpc. Only five $\Theta_0$ measurements (four of them discussed in G17 and one from Salucci~\cite{Salucci13}) were not scaled, as they were independent of $R_0$. \rthis{We note that for their analysis  estimates oc entral values, G17 used only the  post-1985 measurements.}

\section{Analysis}
\label{sec:analysis}

\subsection{Error Distribution and distribution functions}

Similar to Ratra et al (~\cite{Ratra15}), we calculated the estimates using two methods: median statistics  and weighted mean estimates.  The weighted mean ($\Theta_M$) is given by~\cite{Bevington}:
\begin{equation}
\Theta_M = \frac{\sum \limits_{i=1}^N \Theta_i/\sigma_i^2}{\sum \limits_{i=1}^N 1/\sigma_i^2},
\end{equation}
\noindent where $\Theta_i$ indicates each measurement of the rotation and  $\sigma_i$ indicates the total error in each measurement. The total weighted mean error is given by $\sigma_M = \frac{1}{\sum \limits_{i=1}^N 1/\sigma_i^2}$. The goodness of fit is parameterized by $\chi^2$, which is given by 
\begin{equation}
\chi^2= \frac{1}{N-1}\sum \limits_{i=1}^N (\Theta_i-\Theta_m)/\sigma_i^2
\label{eq:chisq}
\end{equation}
The $\chi^2$ defined in Eq.~\ref{eq:chisq} is also referred to as reduced $\chi^2$. 
For a reasonably good fit, this $\chi^2$ has to be close to 1. 

An alternate method to determining the central estimate is using median statistics.
The main advantage of  median statistics-based estimate is that it is robust against outliers and does not make use of individual error bars. \rthis{The median estimate is also expected to be more robust if the error bars are not Gaussian~\citep{Gott,Chen,Bethapudi}.}
More details on  median statistics-based estimates and its applications to a various  astrophysical datasets are reviewed in Refs~\cite{Gott,Chen03,Bethapudi,Chen,Crandall} and references therein. Using the dataset constructed by G17, we calculate the median central estimate of $\Theta_0$ and its 68\% confidence intervals, using the same prescription as in Ref.~\cite{Chen}.
The weighted mean estimate is found to be \rthis{$\Theta_{Mean} = 227.07 \pm 0.70$ km/sec} and the median estimate is calculated to be \rthis{$\Theta_{Med} = 237.58\pm 3.24$ km/sec}. We note that one difference between these estimates and those in G17, is that G17 did the calculations for only the post-1985 measurements, whereas we have included all the measurements tabulated in G17.

For both the estimates of $\Theta_0$, we calculate  $N_{\sigma_i}$, where 
\begin{equation}
N_{\sigma_i} =\frac{\Theta_i-\Theta_{CE}}{\sqrt{\sigma_i^2+\sigma_{CE}^2}}.
\label{eq:nsigma}
\end{equation}

\noindent In the above equation, $\sigma_{CE}$ is the error in the central estimate (which could be the median or weighted-mean based estimate). We now fit the histogram of $N_{\sigma_i}$ to various probability distributions as described in the next section.

\subsection{Fits to probability distributions}

\begin{figure*}
\centering
\includegraphics[width=\textwidth,height=11.5 cm]{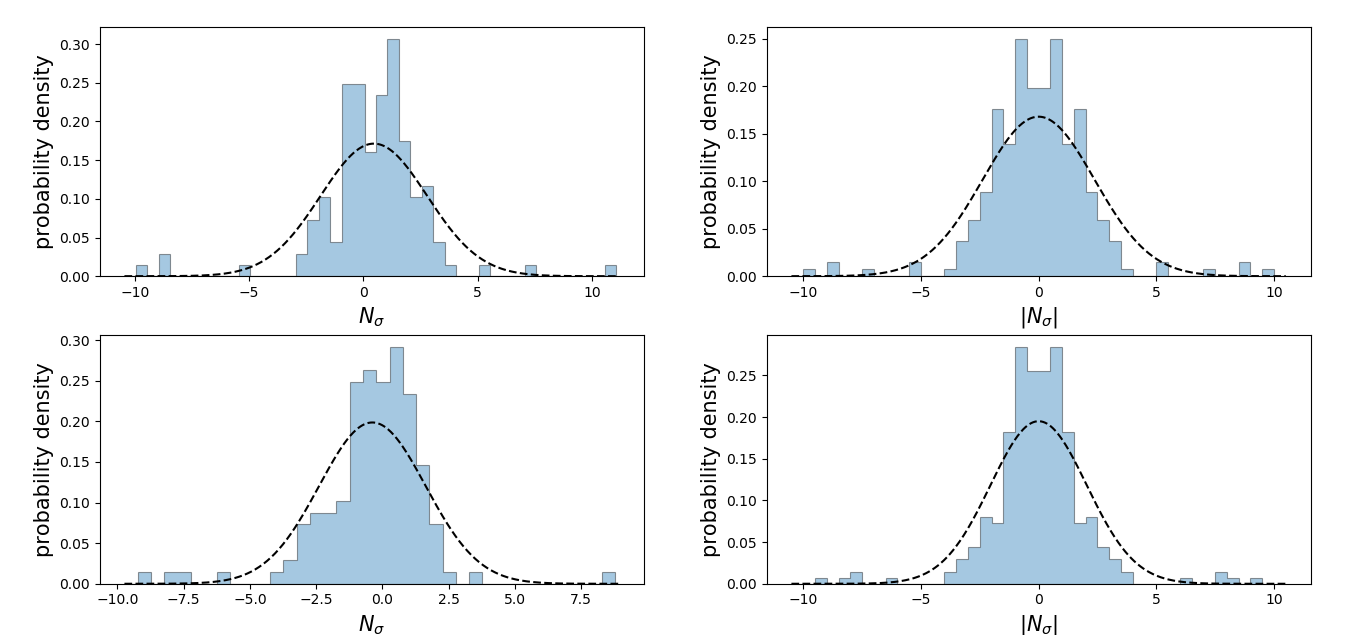}
  \caption{Histograms of the error distributions in half standard deviation bins. The top (bottom) row uses the weighted mean (median) of the \rthis{137} measurements as the central estimate. The left (right) column shows the signed (absolute) deviation. In the left column plots, positive (negative) $N_{\sigma}$ represent a value that is greater (less) than the central estimate. The dotted-line curve is the best fitting Gaussian probability distribution function in all cases. We note that this histograms are normalized so that the sum of the total number of events is equal to 1.}
\label{fig1}
\end{figure*}

We have used $N_{\sigma}$ (defined in Eq.~\ref{eq:nsigma}) for each data point, to construct a histogram for  the error distributions using both the weighted mean and median central estimates. We also construct a corresponding histogram for $|N_{\sigma}|$, where the distributions were symmetrized around the central value, in the same way as in Ref.~\cite{Crandall}. 
These histograms of $|N_{\sigma}|$ and $N_{\sigma}$ are shown in Fig~\ref{fig1}.

\begin{figure}
\centering
\includegraphics[width=0.5\textwidth]{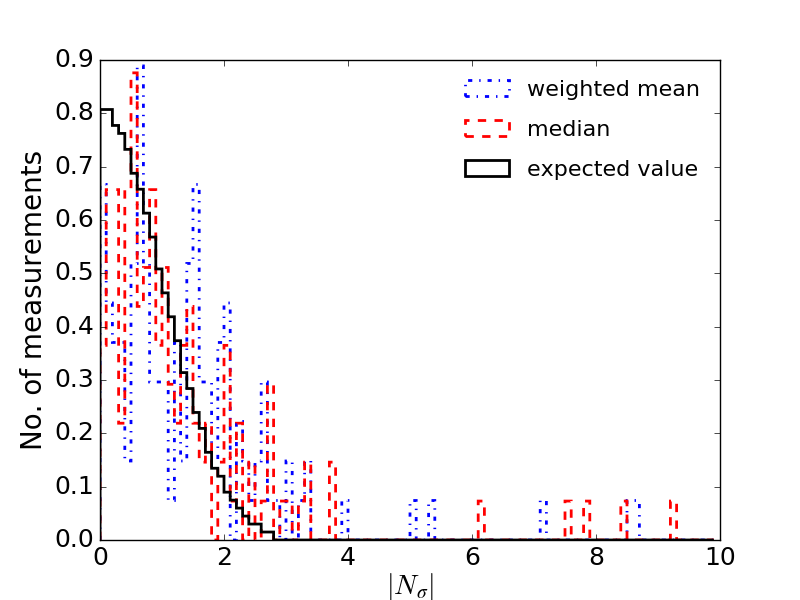}
\caption{ Histogram of the error distributions in $|N_{\sigma}| = 0.1$ bins . The solid black line represents the expected Gaussian probabilities for \rthis{137}
measurements and the dot-dashed blue (dashed red) line is the number of $|N_{\sigma}|$ values in each bin for the weighted mean (median). All the histograms are normalized so that integral of the distribution over all data points is equal to one.}
\label{fig2}
\end{figure}

In terms of $|N_{\sigma}|$, with  mean as the central  estimate, only \rthis{44.53\%} of total distribution lies in the range $\mid N_{\sigma}\mid \leq 1$, and \rthis{75.91\%} of the error distribution lies in the range $\mid N_{\sigma}\mid \leq 2$. When we use the median as the central estimate for the error distribution, \rthis{54.02\%} of the distribution lies in range $|N_{\sigma}| \leq 1$ and \rthis{79.56\%} of the total error distribution lies in the range $|N_{\sigma}| \leq 2$.  However, according to the Gaussian probability distribution, 68.3\% of the total measurements should lie within $\mid N_{\sigma}\mid \leq 1$ and 95.4\% should lie within $\mid N_{\sigma}\mid \leq 2$. Hence, we can conclude that the error distribution in this case deviates from a Gaussian probability distribution to a good extent. \rthis{To further elucidate the discrepancy from normal distribution, we plot in Figure~\ref{fig2} the distribution of $|N_{\sigma}|$ with a bin size of $|N_{\sigma}|$ = 0.1 for both the mean and the median as central estimates compared to the normal distribution. The solid black line shows the expected Gaussian probabilities given by $P(N_{\sigma})= \frac{1}{\sqrt{2\pi}}\exp(-|N_{\sigma}|^2/2)$. The discrepancy from a normal distribution is conspicuous in both the weighted mean and median. We also find a total of five outliers with $|N_{\sigma}|>7$. } 

So, as our next step, we have taken a few well-known non-Gaussian probability distribution functions into consideration, such as Cauchy, Double exponential, and Student-t distributions, in order  to ascertain the probability distribution function which fits well to our error distribution. 

\begin{figure*}
\centering
\includegraphics[width=1.05\textwidth,height=15 cm]{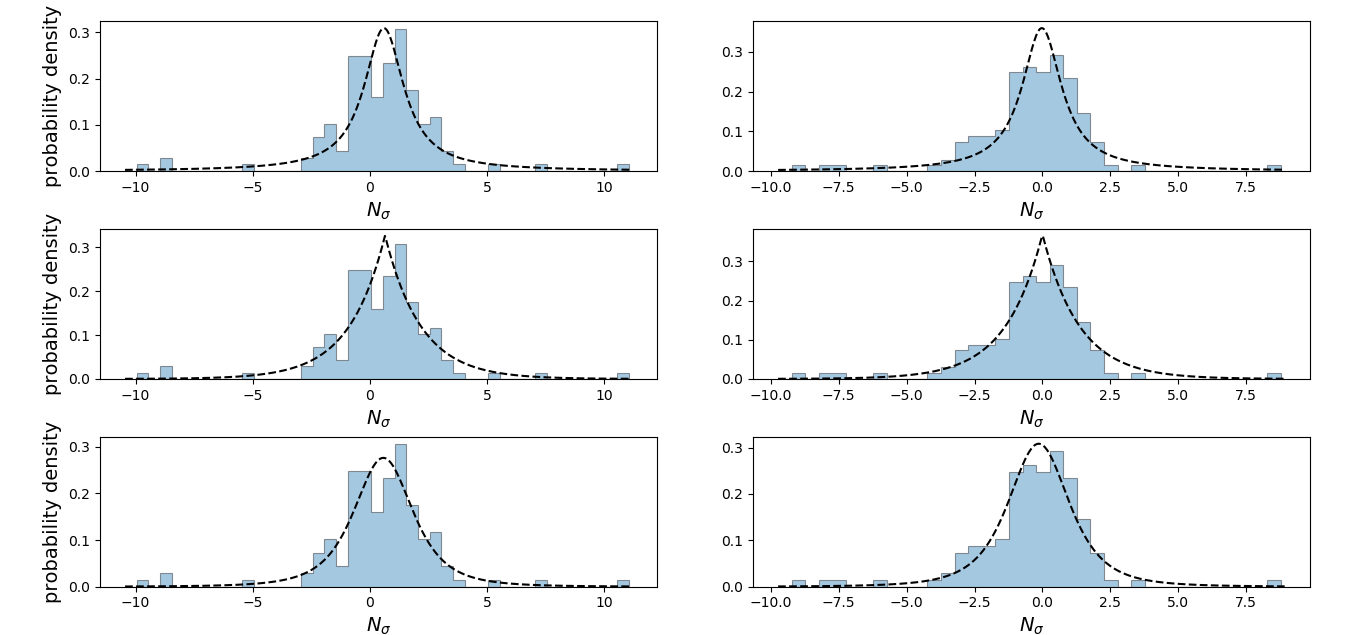}
\caption{The left (right) plot in the each row denotes the histogram for the error distributions with mean (median) as the central estimate. The dotted line curve in the top, middle and bottom rows are the best fitting Cauchy, Double Exponential, and Student's t probability distribution functions respectively. In each case, we  consider mean (median) as the central estimate in the left (right) column.}
\label{fig3}
\end{figure*}

We therefore fit the histograms of $N_{\sigma}$ and  $|N_{\sigma}|$ to four different distributions: Gaussian, Cauchy, double-exponential, and Students-t distribution. We performed the fit using the {\tt stats.fit} functionality in {\tt Scipy} for each of these probability distribution functions. \rthis{Figure~\ref{fig3} shows the Cauchy,  Double exponential and Students-t probability distribution functions, fitted  to the $N_{\sigma}$ histograms using  both the mean and median as the  central estimates.}   
To evaluate the best fit,  we use the distribution-free Kolmogorov-Smirnoff (K-S) test~\cite{NR}. We note that the Students-t distribution has an additional free parameter $n$. We varied the values of $n$ for this distribution until the $p$-value was maximized.  
The K-S probabilities for these distributions both with and without binning, and also with weighted  mean/median as central estimates  are shown in Table~\ref{tab:table1}.

From the corresponding $p$-values for each of the distribution functions, it can be concluded that the Student's t distribution function (with $n=2$) fits comparatively better to our error distribution, when median is taken as the central estimate.  All the remaining distribution functions give  very poor fits. 

\begin{table*}
\begin{center}
\begin{tabular}[t]{|l |c| c| c| c|}
\hline
Function & \multicolumn{2}{|c|}{Un-binned probability} & \multicolumn{2}{|c|}{Binned probability} \\ \hline
  & mean & median  & mean & median\\
Gaussian & \num{8.12e-9} & 0 .0084 & \num{6.13e-12} & 0.0001\\
Cauchy & \num{9.68e-5} & 0.05 &\num{4.89e-7} & 0.022\\
Double exponential & \num{7.84e-8} &0.016 & \num{5.63e-11}  & 0.0005\\
Students-t & (n=1)\num{9.68e-5} & (n=2) 0.1 & (n=1)\num{4.89e-7} & \rthis{(n=1) 0.022}\\ \hline
\end{tabular}
\end{center}
\caption{K-S test probability for various functional fits to $N_{\sigma}$ reconstructed from the  rotation velocity data obtained from the compilation in G17. As we can see, only the Students-t distribution provides a reasonable $p-$ value when median estimate of $\Theta_0$ is used.}
\label{tab:table1}
\end{table*}

\begin{table*}
\begin{center}
\begin{tabular}[t]{|l |c| c| c| c|c|}
\hline
Tracers & No. of Observations* &   Median (km/s)   & 68 \% c.l. (km/s) & p-value (median) & p-value (mean) \\ \hline
 Field stars & 30 &   223.99   & 3.53 & 0.23 & 0.15 \\
Young tracers & 64 & 245.48 & 2.49 & 0.62 & 0.13 \\
Galactic Mass Modeling & 14 & 214.82 & 5.74 & 0.57 & 0.44 \\
Intermediate/old age tracers & 6 & 202.89  & 49.08 & 0.58 & 0.95 \\
SgrA & 10 & 244.46 & 4.72 & 0.11 & 0.21 \\
Others & 12 & 215.31 & 1.95 & 0.17 & 0.28 \\ \hline
\end{tabular}
\end{center}
\flushleft*Observations which have null values for errors are omitted.
\caption{ Median and 68.3\% confidence interval around median for various tracer distributions along with the K-S test probability for Gaussian distribution fit (considering both mean and median as central estimate) for $N_{\sigma}$ reconstructed from the  rotation velocity data obtained from the compilation in G17.}
\label{tab:table2}
\end{table*}

\subsection{Examination of outliers}
\label{sec:outliers}
\rthis{We now briefly discuss some of the measurements, which are the cause of outliers in the $|N_{\sigma}|$ distribution in 
Fig.~\ref{fig2}, to see if a simple explanation can be found for these. When considering the weighted mean, we have five such measurements with $|N_{\sigma}|>7$, corresponding to   $N_{\sigma}$  values of 10.7~\cite{Branham},   -9.7~\citep{Clemens}, -8.6~\citep{Robinson}, -8.4~\cite{Alvarez}, 7.3~\cite{Sharma}.}

\rthis{The largest outlier comes from Branham~\citep{Branham}, corresponding to $\Theta_0=298\pm 7$ km/sec at $R_0$=8.15 kpc. This measurement has been made  using the kinematics of OB stars using data from the Hipparcos satellite  and  by solving for 14 unknowns~\citep{Branham}. Their value of 298 km/sec was obtained after positing a linear model to simplify the equations. When a non-linear model is used, then a value close to our central estimate is obtained.
For the outlier at 9.7$\sigma$~\cite{Clemens}, the website in G17 reports a measurement of 252 $\pm$ 2 km/s for $R_0=10$ kpc. This paper reports a measurement of galactic rotation curve as a function of galacto-centric distance from the UMASS StonyBrook CO survey. However, on closer examination of this paper, 
we find that the galactic rotation curve  was plotted for two different values of $\Theta_0$ and $R_0$, of which one is the value documented by G17. We note however, that no independent estimate of $\Theta_0$ has been  made from the observations. Two ad-hoc values for $\Theta_0$ of 252 and 220 km/sec have been assumed for obtaining the galactic rotation curves. Therefore, this measurement should have been omitted from the database compiled by G17. The next outlier  is at $\Theta_0=180 \pm 6$  km/sec~\citep{Robinson}, corresponding to $8.6\sigma$. This measurement comes from the southern galactic plane CO survey carried out using the 4-m Epping telescope. Their value of 180 km/sec is obtained  from the slope of terminal velocity curve as a function of galactic longitude (Fig. 5(a) of Ref.~\cite{Robinson}). However, this plot contains many outliers and no detailed statistical analysis of the goodness of fit has been made. So there is no guarantee the estimated $\Theta_0$ provided a good fit to the data. The next outlier (at -8.4$\sigma$) is also from CO observations, as part  of the  Deep CO survey of the southern Milky Way galaxy survey~\cite{Alvarez}, which estimated  a value of $\Theta_0= 209 \pm 2 $ km/sec.  One assumption made for this work   was that the galactic rotation curve is completely flat and there is no variation with galactic longitude. The last outlier corresponds to $\Theta_0= 232 \pm 1.7$ km/sec ($N_{\sigma}=7.3$)~\cite{Sharma}. This value was obtained from the Radial Velocity Experiment (RAVE) stellar survey using the Shu distribution function. Although it is hard to discern a specific reason for this high value, one possibility could be that there is a degeneracy between the value of $\Theta_0$ and another parameter defined as $\alpha_z$, which denotes the vertical dependence of the circular speed. The value they obtained for $\alpha_z$ disagrees with  the value of 0.0347 predicted by analytical models of Milky way potential. For smaller values of $\alpha_z$, the estimated values of $\Theta_0$ would also decrease.}

\rthis{Therefore to summarize, we find that one outlier measurement is a consequence of an incorrect tabulation. Two others come from CO  measurements.
The possible reason for the outliers in the other two measurements is a consequence of a simplified model been used in the fitting procedure or due to degeneracy between the rotation speed and another astrophysical parameter.}

\section{Analysis of subsamples}
\label{sec:subsamples}

\rthis{In order to understand the underlying cause of non-Gaussianity when analyzing the full dataset,
we undertake a similar analysis on each subsample of data after grouping the measurements according to  the method used. This will help us determine if there are unknown systematic errors within each group. The classification of $\Theta_0$ measurements has already been done in G17, who divided  the measurements according to the stellar population tracer used. The entries were grouped into field stars; young tracer populations;  old and intermediate age tracers; kinematics of Sgr A* near the galactic center; and  galactic mass modeling  using H1 as well as CO radio observations. 
In G17, central estimates using the weighted mean values of each group has already been calculated, including a discussion of which of these deviate from the IAU recommended values. G17 have found differences among the $\Theta_0$ and $\Theta_0/R_0$ values between the different tracer populations.  They have found that young tracers and kinematic measurements of Sgr A* near the galactic center imply a significantly larger rotation speed at the solar circle compared to the field stars and HI/CO measurements.
Here, we examine the non-gaussianity of errors in each subset to see if there is any underlying unaccounted systematics in each subset of measurements.}

\rthis{For each subset, we carry out the same analysis as in Sect.~\ref{sec:analysis}. We obtain the central estimate using both the weighted mean and median and then construct $N_{\sigma}$ histograms using each of these and fit these to a Gaussian distribution. We check for Gaussianity using the p-values resulting from K-S test. The results can be found in Table~\ref{tab:table2}. To complement the analysis in G17, we show the group-wise medians along with 60\% c.l. ranges obtained using the method by Chen and Ratra~\cite{Chen}. We can see that our median-based estimates for each tracer population agree with the weighted means by G17, except for the intermediate and old age tracer population,
for which we get a value 10 km/sec more than the one by G17.}

\rthis{From Table~\ref{tab:table2}, we find that the p-value for a Gaussian distribution fitting the data is greater than 0.1 for all the subsets, using both the mean as well as the central estimates. Therefore, there is no unknown systematic error or egregious measurement within each group of measurements. The underlying cause of non-gaussianity  for the full dataset is probably caused by combining the data across the tracers, in addition to outliers.}

\rthis{Finally, similar to previous works on median statistics estimates of astrophysical and cosmological parameters, we obtain a central estimate by calculating median of this group-wise median estimates. This central estimate from the median of all these medians is given by $\Theta_0$=219.65 km/sec. The total number of measurement categories is too small to get a robust 68\% confidence level error bar on this value.}
 
\rthis{Given the non-Gaussianity of the residuals from the full dataset, this median value of 219.65 km/sec should be used as the central estimate of $\Theta_0$. We note that this value is closer to the IAU recommended value of 220 km/sec and differs from the recommendation in G17 of 225 km/sec inspite of using the same galactocentric distance of 8.3 kpc.}

\section{Conclusions}
\label{sec:conclusions}
We have used a compilation of 137 measurements of Galactic rotation speed and their corresponding errors from G17~\citep{Grijs} and two additional measurements (not included in G17), in order to gain a better insight of the non-Gaussianity of the residuals \rthis{and to obtain a central estimate}. We first scaled all the measurements, which were  degenerate with galactocentric distance  to   a common  value of 8.3 kpc. The error distributions were analyzed (following the same prescription as in the previous works by Ratra et al~\cite{Ratra03,Ratra15,RatraLMC,RatraGC}) and plotted using both the weighted mean  as well as the median value   as the central estimate. We note that the central estimates for the weighted mean and median used all the measurements unlike those in G17, which used only the post-1985 measurements.

We conclude from our observations that the error distribution for the galactic rotation speed measurements using both these estimates is inherently non-Gaussian.
The deviation from Gaussian distribution motivated us to check the fit for other prominent non-Gaussian probability distribution functions. We have taken into consideration the Cauchy, Double exponential and Students-t probability distribution functions.  The results show that with median as the central estimate, the error distribution have comparatively better fits with Student's t probability distribution functions  for $n=2$, corresponding to a  $p$-value of 0.1. All other distributions display poor fits with both mean and median values as central estimates.

\rthis{We then redid the same analysis   after grouping the measurements according to the tracers used. We find  that the residuals within each subsample follow the Gaussian distribution. This implies that the non-Gaussianity of the error bars is caused by combining the measurements from different categories, in addition to outliers.}

\rthis{Finally, since the residuals are not Gaussian, instead of the weighted mean, the median value when grouped according to the  measurement type should be used as the central estimate for $\Theta_0$. This group-wise median value is equal to 219.65 km/sec and is close to the IAU recommended value of 220 km/sec. This is inspite of using a galactocentric distance of 8.3 kpc.}

\begin{acknowledgements}
\rthis{We thank the anonymous referee for invaluable
constructive feedback on the manuscript draft. We are also grateful to Richard De Grijs and Paolo Salucci for useful correspondence.}
\end{acknowledgements}

\bibliography{main}
\end{document}